\documentclass[conference]{IEEEtran}
\usepackage{cite}
\usepackage{amsfonts,amsbsy,amsmath}
\usepackage{amsthm}
\usepackage{amssymb}
\usepackage{dsfont}
\usepackage{enumitem} 
\usepackage{tikz}
\usepackage{pgfplots}
\usepackage[detect-all,binary-units]{siunitx}
\usepackage{bm,bbm}
\usepackage{booktabs}
\usepackage{multirow}
\usepackage{color}
\usepackage[caption=false,font=footnotesize]{subfig}
\usepackage{mathtools}
\mathtoolsset{showonlyrefs}
\usepackage{mathrsfs}
\usepackage{graphicx}
\usepackage{scalerel}
\usepackage{pgfplots}
\usepackage{balance}
\usepackage[nolist]{acronym}

\theoremstyle{plain}

\newtheorem{lemma}{Lemma}

\allowdisplaybreaks

\renewcommand{\Pr}[1]{\mathds{P}\!\left(#1\right)}

\newcommand{\ldeg}{\mathtt{l}}
\newcommand{\rdeg}{\mathtt{r}}
\newcommand{\bldeg}{\bar{\ldeg}}
\newcommand{\brdeg}{\bar{\rdeg}}

\newcommand{\Cirreg}{\mathscr{P}^{(n)}_{\Lambda,\Rho}}
\newcommand{\Creg}{\mathscr{P}^{(n)}_{\ldeg,\rdeg}}
\newcommand{\Cregn}[3]{\mathscr{P}^{(#1)}_{#2,#3}}

\newcommand{\vn}{\mathtt{v}}
\newcommand{\cn}{\mathtt{c}}

\newcommand{\cN}{\mathcal{N}}
\newcommand{\cV}{\mathcal{V}}

\newcommand{\cC}{\mathcal{C}}
\newcommand{\cE}{\mathcal{E}}

\newcommand{\coeff}[1]{\mathrm{coeff}\!\left[#1\right]}

\newcommand{\algD}{\scaleto{\mathsf{D}}{4pt}}
\newcommand{\MD}{\scaleto{\mathsf{MD}}{4pt}}
\newcommand{\FA}{\scaleto{\mathsf{FA}}{4pt}}
\newcommand{\DD}{\scaleto{\mathsf{DD}}{4pt}}
\newcommand{\COMP}{\scaleto{\mathsf{COMP}}{4pt}}

\newcommand{\pfa}{P_{\FA}}
\newcommand{\pfn}{P_{\MD}}

\newcommand{\dumx}{\mathtt{x}}
\newcommand{\dumy}{\mathtt{y}}
\newcommand{\dums}{\mathtt{s}}
\newcommand{\dumt}{\mathtt{t}}

\newcommand{\LA}{\Lambda}
\newcommand{\RH}{\mathrm{P}}
\newcommand{\Rho}{\mathrm{P}}

\newcommand{\defset}{\mathcal{I}}
\newcommand{\hdefset}{\hat{\mathcal{I}}}
\newcommand{\hstate}{\hat{x}}
\newcommand{\state}{x}

\definecolor{TUMBlack}{cmyk}{0,0,0,1}     
\definecolor{TUMWhite}{cmyk}{0,0,0,0}     

\definecolor{TUMBlue} {cmyk}{1,0.43,0,0}  

\definecolor{TUMDarkBlue}   {cmyk}{1,0.57,0.12,0.7}      
\definecolor{TUMDarkerBlue} {cmyk}{1,0.54,0.04,0.19}     
\definecolor{TUMMediumBlue} {cmyk}{0.9,0.48,0,0}         
\definecolor{TUMLighterBlue}{cmyk}{0.65,0.19,0.01,0.04}  
\definecolor{TUMLightBlue}  {cmyk}{0.42,0.09,0,0}        

\definecolor{TUMDarkGray}  {cmyk}{0,0,0,0.8}  
\definecolor{TUMMediumGray}{cmyk}{0,0,0,0.5}  
\definecolor{TUMLightGray} {cmyk}{0,0,0,0.2}  

\definecolor{TUMGreen} {cmyk}{0.35,0,1,0.2}         
\definecolor{TUMOrange}{cmyk}{0,0.65,0.95,0}        
\definecolor{TUMIvory} {cmyk}{0.03,0.04,0.14,0.08}  

\definecolor{TUMBeamerYellow}    {rgb}{1.00,0.71,0.00}  
\definecolor{TUMBeamerOrange}    {rgb}{1.00,0.50,0.00}  
\definecolor{TUMBeamerRed}       {rgb}{0.90,0.20,0.09}  
\definecolor{TUMBeamerDarkRed}   {rgb}{0.79,0.13,0.25}  
\definecolor{TUMBeamerBlue}      {rgb}{0.00,0.60,1.00}  
\definecolor{TUMBeamerLightBlue} {rgb}{0.25,0.75,1.00}  
\definecolor{TUMBeamerGreen}     {rgb}{0.57,0.67,0.42}  
\definecolor{TUMBeamerLightGreen}{rgb}{0.71,0.79,0.51}  

\definecolor{I8LogoRed}         {rgb}{0.51,0,0.08}
\definecolor{I8LightBlue}       {rgb}{0.725,0.812,0.882}

\newcommand{\bttx}{\text{\texttt{\textbf{x}}}}
\newcommand{\bttt}{\text{\texttt{\textbf{t}}}}
\newcommand{\btts}{\text{\texttt{\textbf{s}}}}

\IEEEoverridecommandlockouts

\begin{document}

\title{Ensemble Average Analysis of Non-Adaptive Group Testing with Sparse Pooling Graphs}

\author{\IEEEauthorblockN{Emna Ben Yacoub$^{\ddagger}$, Gianluigi Liva$^\dagger$, Enrico Paolini$^*$, Marco Chiani$^*$}
	\IEEEauthorblockA{$^\ddagger$Huawei Munich Research Center, Munich, Germany\\$^\dagger$Institute of Communications and Navigation, German Aerospace Center (DLR), Germany\\$^*$University of Bologna,
		Cesena, Italy
}}

\maketitle
\thispagestyle{empty}

\begin{abstract}
	A combinatorial analysis of the false alarm (FA) and misdetection (MD) probabilities of non-adaptive group testing with sparse pooling graphs is developed. The analysis targets the combinatorial orthogonal matching pursuit and definite defective detection algorithms in the noiseless, non-quantitative setting. The approach follows an ensemble average perspective, where average FA/MD probabilities are computed for pooling graph ensembles with prescribed degree distributions. The accuracy of the analysis is demonstrated through numerical examples, showing that the proposed technique can be used to characterize the performance of non-adaptive group testing schemes based on sparse pooling graphs.
\end{abstract}

\thispagestyle{empty}
\setcounter{page}{1}

\begin{acronym}
	\acro{FA}{false alarm}
	\acro{MD}{misdetection}
	\acro{BP}{belief propagation}
	\acro{MAP}{maximum a posteriori}
	\acro{CDF}{cumulative distribution function}
	\acro{PDF}{probability density function}
	\acro{LDPC}{low-density parity-check}
	\acro{FER}{frame error rate}
	\acro{SER}{symbol error rate}
	\acro{SPB}{sphere packing bound}
	\acro{DE}{density evolution}
	\acro{MI}{mutual information}
	\acro{LLR}{log-likelihood ratio}
	\acro{i.i.d.}{independent and identically distributed}
	\acro{BER}{bit error rate}
	\acro{DMS}{discrete memoryless source}
	\acro{ML}{maximum likelihood}
	\acro{BSC}{binary symmetric channel}
	\acro{MET}{multi edge type}
	\acro{AS}{absorbing set}
	\acro{EAS}{elementary absorbing set}
	\acro{TS}{trapping set}
	\acro{FAS}{fully absorbing set}
	\acro{EFAS}{elementary fully absorbing set}
	\acro{ETS}{elementary trapping set}
	\acro{DMC}{discrete memoryless channel}
	\acro{SPC}{single parity-check}
	\acro{APP}{a posteriori probability}
	\acro{CVWEF}{composition vector weight enumerator function}
	\acro{WEF}{weight enumerator function}
	\acro{VWEF}{vector weight enumerating function}
	\acro{CCVWEF}{complete composition vector weight enumerator function}
	\acro{SS}{stopping set}
	\acro{COMP}{combinatorial orthogonal matching pursuit}
	\acro{DND}{definitely non-defective}
	\acro{DD}{definite defective}
	\acro{PD}{possible defective}
	\acro{FAR}{false alarm rate}
	\acro{MDR}{misdetection rate}
\end{acronym}

\section{Introduction}\label{sec:intro}

The recent COVID-$19$ pandemic triggered renewed interest in
group testing \cite{dorfman} as a practical solution to the detection of small sets of infected individuals in a large population \cite{mallapaty2020mathematical,narayanan2020pooling,CLP:Identificationdetection.2022}. Among non-adaptive and non-quantitative group testing algorithms \cite{aldridge2019group}, \ac{COMP} \cite{KS64,chan_noisycomp}, and \ac{DD} detection provide simple and effective approaches. In a \ac{COMP} detection, all participants in a negative pool are marked as \emph{non-defective}, and all remaining individuals as \emph{defective}. Therefore, \ac{COMP} achieves a zero \ac{MD} probability, while the \ac{FA} is strictly positive. \ac{DD} works in a dual manner: as \ac{COMP}, it marks all participants in a negative group as non-defective. The remaining individuals are temporarily labeled as \emph{possibly defective}. A possible defective element is marked as defective if and only if it is the only possible defective element in a positive pool. As a result, \ac{DD} achieves a zero \ac{FA} probability, while the \ac{MD} is strictly positive.

In this paper, we introduce a combinatorial analysis on the \ac{FA} and \ac{MD} probabilities of non-adaptinve group testing with sparse pooling graphs \cite{sejdinovic-johnson,lee2019saffron,NarayananAllerton,wadayama17,KarimiGT} based on \ac{COMP} and \ac{DD} detection.
The analysis focuses on the noiseless case and is based on generating functions to enumerate configurations of binary labels assigned by the detector to the population elements. Enumeration through generating functions has previously been applied to analyze sparse graph error-correcting codes, e.g., in \cite{BM:Asymptotic.2004a,Orlitsky05,DUR06,FPCF:Spectral.2013,Ben23}. In the context of group testing, the enumeration through generating functions was explored in \cite{wadayama17} to analyze the typical set detection performance. As in \cite{wadayama17}, we embrace an ensemble average perspective, where average FA/MD probabilities are analyzed for pooling graph ensembles with prescribed degree distributions. We provide an exact expression of the ensemble average \ac{FA} probability under \ac{COMP} for a defectiveness Bernoulli model. Similarly, we derive the exact expression of the ensemble average \ac{MD} probability under \ac{DD} detection. 

\section{Preliminaries}\label{sec:prel}

\subsection{Main Definitions}
We consider a noiseless non-adaptive group testing problem in which $m$ tests are performed on a population of $n$ items to identify defective items. We introduce the defectivity vector $\bm{\state} = (\state_1,\state_2,\ldots,\state_n)$ and the corresponding random vector $\bm{X}=(X_1,X_2,\ldots,X_n)$, where $\state_i=1 $ if the $i$-th item is defective and $\state_i=0$ otherwise. We consider an \ac{i.i.d.}  Bernoulli model, where the prevalence is $\Pr{X_i=1}=\delta$. Consider a generic detection algorithm $\mathsf{D}$, which takes as input the result of $m$ tests and produces the decision on the status of the elements $\hat{\bm{\state}}$. Denoting by $\hat{\bm{X}}$ the corresponding random vector, the \ac{FA} and the \ac{MD} probabilities under $\mathsf{D}$ are 
\begin{align}
	\pfa^{\algD}=&\frac{1}{n}\sum_{\ell=1}^{n}\Pr{\hat{X}_\ell=1 | X_\ell=0} \\
	\pfn^{\algD}=&\frac{1}{n}\sum_{\ell=1}^{n}\Pr{\hat{X}_\ell=0 | X_\ell=1}.
\end{align}

\textit{Notation}: For a multivariate polynomial $p(\bttx)$, where $\bttx=(\dumx_1,\dumx_2,\ldots,\dumx_k)$, we denote by $\coeff{p(\bttx),\dumx_1^{i_1}\dumx_2^{i_2}\cdots\dumx_k^{i_k}}$ the coefficient of $\prod_{\ell=1}^k \dumx_\ell^{i_\ell}$ in $p(\bttx)$.

\smallskip

\textit{Pooling Graphs}: Let $G= (\cV, \cC, \cE)$ be a bipartite graph, called a pooling graph, consisting of a set $\cV$ of $n$ left nodes (one for each element),  a set $\cC$ of $m<n$ right nodes (one for each test) and a set of edges $\cE=\{e_{ij}\}$, where $e_{ij}$ is an edge between the left node $\vn_j\in\cV$ and the right node $\cn_i\in\cC$.
Let $\defset$ be the set of left nodes corresponding to defective items and $\hdefset$ the estimate of $\defset$. We denote by $(\state_i,\hstate_i)$ the state of the left node $\vn_i$. We have $\state_i =1$ if $\vn_i \in \defset$, $\state_i=0$ otherwise, and $\hstate_i=1$ if $\vn_i \in \hdefset$,  and $\hstate_i=0$ otherwise. An item is false detected if $\state_i \neq \hstate_i$. 
According to the noiseless model, a test on a subset of items is negative if all of these items are non-defective and is positive otherwise.  We say that a right node is positive (negative) if its corresponding test is positive (negative), i.e., a right node is positive if it is connected to at least one left node in $\defset$ and negative otherwise.
The sets $\cN(\vn)$ and $\cN(\cn)$ denote the neighbors of $\vn$ and $\cn$. The degree of a left node $\vn$ (right node $\cn$)    is the cardinality of the set $\cN(\vn)$ ($\cN(\cn)$).
The node-oriented degree distribution polynomials of a pooling graph are  
$\Lambda(x)=\sum_{i}\Lambda_{i}x^{i}$ and $
\Rho \left(x\right) = \sum_{i}\Rho_{i}x^{i}$
where $ \Lambda_{i} $ is the fraction of left nodes with degree $ i $ and $ \Rho_{i} $ is the fraction of right nodes with degree $ i$.  We denote by 
$\bldeg = \sum_{i} i \Lambda_i$ and $\brdeg  = \sum_{i} i \Rho_i$
the average left node and right node degrees, respectively. Note that $n \bldeg = m \brdeg$ represents the total number of edges. The minimum (maximum) left node degree is $\ldeg_{\text{min}}$ ($\ldeg_{\text{max}}$). Similarly, the minimum (maximum) right node degree is $\rdeg_{\text{min}}$ ($\rdeg_{\text{max}}$).
We denote the ratio between the number of tests and the number of elements as $\xi = m/n = \bldeg/\brdeg <1$.
An irregular pooling graph ensemble $\Cirreg$ is the set of all graphs with  $ n $ left nodes and degree distributions $ \Lambda\left( x\right) $ and $ \Rho\left( x\right) $.
A pooling graph is called $(\ldeg, \rdeg)$-regular if all left nodes have the same degree $\ldeg$ and all right nodes have the same degree $\rdeg$. We denote by $\Creg$  the set of all pooling graphs with $ n $
left nodes,  left node degree $\ldeg$ and right node degree $\rdeg$.

\smallskip

\textit{COMP Detection Algorithm}:
The \ac{COMP} algorithm is based on the fact that any element in a negative test must be  \ac{DND}, and it assumes that the other elements are defective. This means that we mark any left node connected to at least one negative right node as \ac{DND}, and label the remaining left nodes  \ac{PD}. Then $\hdefset_{\COMP} $ is the set of \ac{PD} items. The \ac{COMP} algorithm does not generate false negatives (misdetections).

\smallskip

\textit{DD Detection Algorithm}:
The \ac{DD}  algorithm uses the observation that if a positive test contains exactly one \ac{PD}, then that item is actually \ac{DD}. Similarly to \ac{COMP}, a left node connected to at least one negative right node is declared as \ac{DND}. The remaining left nodes are called \ac{PD}.
If a \ac{PD} left node is the only \ac{PD} connected to a positive right node, we call it \ac{DD}. The set of \ac{DD} elements is $\hdefset_{\DD} $. The \ac{DD} algorithm does not produce false positives (false alarms).

\section{COMP Detection Analysis}\label{sec:COMP}
In this section, we derive the average number of patterns with $i$ defective elements and $j$ false detected candidates under \ac{COMP} for regular and irregular ensembles. The enumerators are then used to compute the ensemble average \ac{FA} probability.

\begin{lemma}[Average enumerator for regular ensembles]\label{lemma:ACOMPreg}
	Under \ac{COMP} detection, the average number of patterns with $i$ defective elements and $j$ false detected candidates in a pooling graph drawn uniformly at random from the regular ensemble $\Creg$ is given by
	\begin{equation}\label{eq:ACOMPreg}
		\begin{aligned}
			A_{i,j}^{\COMP}=& {n \choose i,j,n-i-j} \sum\limits_{b=\lceil (i+j)\ldeg/\rdeg \rceil }^{m}  {m \choose b} \times \\&\frac{  \coeff{g(\dumx,\dumy)^b,\dumx^{j \ldeg}\dumy^{b \rdeg-\ldeg(i+j)}} }{{n\ldeg \choose \ldeg i,  \ldeg j,b \rdeg-\ldeg(i+j), (m-b)\rdeg}} \times \\&  \coeff{f(\dums)^{n-i-j},\dums^{b \rdeg-\ldeg (i+j)}}  
		\end{aligned}
	\end{equation}
	where $b$ represents the number of positive tests and 
	\begin{align} 
		g(\dumx, \dumy)=&(1+\dumx+\dumy)^\rdeg -(\dumx+\dumy)^\rdeg \label{eq:gCOMPreg}\\
		f(\dums)=&(1+\dums)^\ldeg -\dums^\ldeg. \label{eq:fCOMPreg}
	\end{align}
	\begin{proof}
		Consider a random graph in $\Creg$. We randomly chose a set $\defset$ of $i$ left nodes ($\defset$ is the set of defectives). 
		We have three types of edges emanating from positive right nodes. Edges of the first type are connected to left nodes with state $(1,1)$, edges of the second type are connected to left nodes with state $(0,1)$, and edges of the third type are connected to left nodes with state $(0,0)$. Note that nodes with state $(1,0)$ are impossible since \ac{COMP} does not produce false negatives. Observe that there are $n-i-j$  left nodes with state $(0,0)$ and each one of them is connected to at least one negative right node. Let $b$ be the number of positive right nodes. Each positive right node is connected to at least one left node in $\defset$.
		The number of configurations with $\ldeg i$ type-$1$ edges,  $\ldeg j$ type-$2$ edges, $b \rdeg - \ldeg (i+j)$ type-$3$ edges, $i$ left nodes with state $(1,1)$ and  $j$ left nodes with state $(0,1)$ connected each to $\ldeg$ positive right nodes, and $n-i-j$ left nodes with state $(0,0)$, is
		\begin{equation}\label{eq:alphaCOMPreg}
			{n \choose i,j,n-i-j}  \coeff{f(\dums)^{n-i-j},\dums^{b \rdeg-\ldeg(i+j)}}
		\end{equation}
		where $f(\dums)$ is defined in \eqref{eq:fCOMPreg}.
		Moreover, the number of configurations with $\ldeg  i$ type-$1$ edges, $\ldeg  j$ type-$2$ edges, and $b \rdeg-\ldeg(i+j)$ type-$3$ edges where there are exactly $b$ positive right nodes, each connected to at least one left node with state $(1,1)$, is
		\begin{equation}\label{eq:betaCOMPreg}
			{m \choose b} \coeff{g(\dumx,\dumy)^b,\dumx^{j \ldeg}\dumy^{b \rdeg-\ldeg(i+j)}}
		\end{equation}
		where $g(\dumx,\dumy)$ is defined in \eqref{eq:gCOMPreg}. 
		The proof follows by multiplying \eqref{eq:alphaCOMPreg} by \eqref{eq:betaCOMPreg}, dividing the result by the total number of configurations for the given distribution of edge types, and summing over all possible $b$.
	\end{proof}
\end{lemma}

\begin{lemma}[Average enumerator for irregular ensembles]
	Under \ac{COMP} detection, the average number of patterns with $i$ defective elements and $j$ false detected candidates in a pooling graph drawn uniformly at random from the irregular ensemble $\Cirreg$ is given by
	\begin{equation}
		\begin{aligned}
			A_{i,j}^{\COMP}=&  \sum\limits_{w,e,u} \frac{  \coeff{g(\bttx)^n,\dumx_1^{w}\dumx_2^{e} \dumx_3^{u}}   }{{n{\bldeg} \choose w,e,u,n \bldeg-w-e-u}} \times  \\& \coeff{f(\bttt,\btts)^{n},\dumt_1^i   \dumt_2^j    \dums_1^w \dums_2^e \dums_3^u}
		\end{aligned}
	\end{equation}
	where $\bttx=(\dumx_1,\dumx_2,\dumx_3)$, $\bttt=(\dumt_1,\dumt_2)$, $\btts=(\dums_1,\dums_2,\dums_3)$,
	\begin{align}
		g(\bttx)=&\prod\limits_{d=1}^{\rdeg_{\text{max}}} \left[ 1+ (\dumx_1+\dumx_2+\dumx_3)^d- (\dumx_2+\dumx_3)^d \right]^{\xi \RH_d}\\
		f(\bttt,\btts)=&\prod\limits_{d=1}^{\ldeg_{\text{max}}} \left[ \dumt_1 \dums_1^d+ \dumt_2 \dums_2^d+(1+\dums_3)^d-\dums_3^d \right]^{ \LA_d}
	\end{align}
	and the sum is over all integers $w,e,u$ such that
	$i \ldeg_{\text{min}} \leq w \leq i \ldeg_{\text{max}}$, $ 
	j \ldeg_{\text{min}} \leq e \leq j \ldeg_{\text{max}}$, $0 \leq u \leq (n-i-j) (\ldeg_{\text{max}}-1)$, and $
	w+e+u \leq n \bldeg$.
	\begin{proof}
		Consider a pooling graph drawn uniformly at random from the ensemble $\Cirreg$. We randomly chose a set $\defset$ of $i$ left nodes ($\defset$ is the set of defectives). 
		Here again  we have three types of edges emanating from positive right nodes. Edges of the first type are connected to left nodes with state $(1,1)$, edges of the second type are connected to left nodes with state $(0,1)$ and edges of the third type are connected to left nodes with state $(0,0)$. We denote by $\alpha(i,j,w,e,u)$ the  number of ways to choose $i$ defective left nodes (state $(1,1)$) such that we have  exactly $j$ left nodes with state $(0,1)$,  $w$ type-$1$ edges, $e$ type-$2$ edges and $u$ type-$3$ edges. Its generating function is 
		\[
		\sum_{i,j,w,e,u} \alpha(i,j,w,e,u) t_1^i t_2^j  \dums_1^w \dums_2^e \dums_3^u.
		\]
		Consider a  single left node  of degree $d$. If it has the state $(1,1)$, then we obtain $d$ type-$1$ edges.  If it has the state $(0,1)$, then we obtain  $d$ type-$2$ edges. If it has state $(0,0)$, then it should be connected to at most $d-1$ type-$3$ edges since it should be connected to at least one negative right node (recall that edge types are defined for left nodes connected to at least one positive right node). Considering all possible left node degrees and since we have $n \LA_d$ degree $d$ left nodes, we obtain 
		\begin{equation}\label{eq:alphaCOMPirr}
			\alpha(i,j,w,e,u)= \coeff{f(\bttt,\btts)^n,\dumt_1^i   \dumt_2^j    \dums_1^w \dums_2^e \dums_3^u}.
		\end{equation}
		Let $\beta(w,e,u)$ be the number of ways to choose  $w$ type-$1$ edges, $e$ type-$2$ edges and $u$ type-$3$ edges. Its generating function is given by
		\[
		\sum_{w,e,u} \beta(w,e,u) \dumx_1^w \dumx_2^e \dumx_3^u.
		\]
		Consider a single right node of degree $d$. If it is negative, then we obtain zero type-$1$, $2$, and $3$ edges. If it is positive, then we obtain at least one type-$1$ edge and at most $d-1$ type-$2$ and $3$ edges.
		Considering all possible right node degrees and since we have $m  \RH_d$ degree $d$ right nodes, we obtain 
		\begin{equation}\label{eq:betaCOMPirr}
			\beta(w,e,u)= \coeff{g(\bttx)^n  ,   \dumx_1^w \dumx_2^e \dumx_3^u}.
		\end{equation}
		The proof follows by multiplying \eqref{eq:alphaCOMPirr} with \eqref{eq:betaCOMPirr}, dividing the result by the total number of configurations for the given distribution of edge types, and summing over the edge types distribution $(w,e,u)$.
		\par \vspace{-1\baselineskip}
		\qedhere\end{proof}
\end{lemma}
\subsection{False Alarm Probability}
The probability of having $j$ false alarms conditioned on having $i$ defective elements is $Q(j|i)={n \choose i}^{-1}A_{i,j}^{\COMP}$
and the \ac{FA} probability for a population with prevalence $\delta$ is
\begin{align}
	\pfa^{\COMP}=&\sum\limits_{i=1}^{n}   \sum\limits_{j=1}^{n-i}\frac{j}{n-i} Q(j|i){n \choose i}\delta^i(1-\delta)^{n-i} \\=&\sum\limits_{i=1}^{n} \sum\limits_{j=1}^{n-i}\frac{j}{n-i}A_{i,j}^{\COMP} \delta^i(1-\delta)^{n-i}.\label{eq:pfa}
\end{align}

\section{Definite Defective Detection Analysis}\label{sec:DD}

In this section, we derive the average number of patterns with $i$ defective elements and $j$ falsely detected candidates under \ac{DD} for regular and irregular ensembles. The enumerators are then used to compute the ensemble average \ac{MD} probability.
\begin{lemma}[Average enumerator for the regular ensembles]\label{lemma:ADDreg}
	Under \ac{DD} detection, the average number of patterns with $i+j$ defective elements and $j$ misdetections in a pooling graph drawn uniformly at random from the regular ensemble $\Creg$ is given by
	\begin{equation}\label{eq:ADDreg}
		\begin{aligned}
			&A_{i+j,j}^{\DD}= \sum\limits_{k,b_1,b_2} {n \choose i,j,k,n-i-j-k}  \times  \\& {m \choose b_1,b_2,m-b_1-b_2} \coeff{f_1(\dums_1)^{i},\dums_1^{i \ldeg-b_1}}  \rdeg^{b_1}  \times \\& \coeff{f_2(\dums_2,\dums_3)^{n-i-j-k},\dums_2^{b_2 \rdeg+b_1-(i+j+k)\ldeg} \dums_3^{b_1 (\rdeg-1)}}\times  \\ & \frac{  \coeff{g(\bttx)^{b_2},\dumx_1^{b_2 \rdeg+b_1-(i+j+k)\ldeg}\dumx_2^{k \ldeg} \dumx_3^{i \ldeg -b_1}} }{{n\ldeg \choose (m-b_1-b_2)\rdeg, j \ldeg, b_2 \rdeg+b_1-(i+j+k)\ldeg, b_1 (\rdeg-1), k \ldeg, i \ldeg -b_1, b_1  }}  
		\end{aligned}
	\end{equation}
	where $\bttx=(\dumx_1,\dumx_2,\dumx_3)$, $b_1+b_2$ represents the number of positive tests and the sum is over $k, b_1, b_2$ satisfying 
	$i  \leq b_1 \leq \min\left\{i \ldeg, (n-i-j-k)\ldeg/(\rdeg-1)\right\}$, $
	b_1+b_2 \leq \min\{ m, (i+j)\ldeg\}$, $
	(i+j+k) \ldeg \leq b_2 \rdeg+b_1$, 
	$n-j-k \leq (m-b_1-b_2) \rdeg$, 
	$0 \leq n-i-j-k$,
	and where
	\begin{align} 
		\begin{split} \label{eq:gDDreg}
			g(\bttx)\!=&(1\!+\!\dumx_1\!+\!\dumx_2\!+\!\dumx_3)^\rdeg \!\!-\!(\dumx_1\!+\!\dumx_2)^\rdeg\!-\!\rdeg \dumx_1^{\rdeg-1}\!-\!\rdeg \dumx_3 \dumx_1^{\rdeg-1}
		\end{split}
		\\
		f_1(\dums_1)=&(1+\dums_1)^\ldeg -\dums_1^\ldeg
		\label{eq:f1DDreg} \\
		f_2(\dums_2,\dums_3)=&(1+\dums_2+\dums_3)^\ldeg -(\dums_2+\dums_3)^\ldeg.
		\label{eq:f2DDreg}
	\end{align}
	\begin{proof}
		Consider a pooling graph drawn uniformly at random from the ensemble $\Creg$. We randomly choose a set $\defset$ of $i+j$ left nodes ($\defset$ is the set of defectives).
		We have two types of positive right nodes. Positive right nodes of the first type are connected to exactly one \ac{PD} left node (which will be detected as $(1,1)$). The remaining positive right nodes are positive right nodes of the second type. We denote by $b_1$ the number of type-$1$ positive right nodes and $b_2$ the number of type-$2$ positive right nodes.
		We have six types of edges emanating from positive right nodes. Edges of the first type are connected to left nodes with state $(1,0)$ and type-$2$ positive right nodes, edges of the second type are connected to \ac{DND} left nodes, and type-$2$ positive right nodes, edges of the third type are connected to \ac{DND} left nodes and type-$1$ positive right nodes, edges of the fourth type are connected to negative left nodes that are not connected to any negative right nodes and type-$2$ positive right nodes, edges of the fifth type are connected to \ac{DD}  left nodes, and type-$2$ positive right nodes and edges of the sixth type are connected to \ac{DD}  left nodes and type-$1$ positive right nodes.
		The number of configurations with $\ldeg j$ type-$1$ edges,  $b_2 \rdeg -(i+j+k) \ldeg +b_1$ type-$2$ edges, $b_1 (\rdeg-1)$ type-$3$ edges, $k \ldeg $ type-$4$ edges, $ i \ldeg -b_1$ type-$5$ edges, $b_1$ type-$6$ edges, $i$ \ac{DD} left nodes (state $(1,1)$), $j$ left nodes with state $(1,0)$, $k$ negative left nodes that are not connected to any negative right nodes and $n-i-j-k$ negative left nodes that are connected to at least one negative right node (\ac{DND}),  is
		\begin{equation}\label{eq:alphaDDreg}
			\begin{aligned}
				&{n \choose i,j,k,n-i-j-k} \coeff{f_1(\dums_1)^{i},\dums_1^{i \ldeg-b_1}} \times  \\& \coeff{f_2(\dums_2,\dums_3)^{n-i-j-k},\dums_2^{b_2 \rdeg+b_1-(i+j+k)\ldeg}\dums_3^{b_1 (\rdeg-1)}} 
			\end{aligned}
		\end{equation}
		where $f_1(\dums_1)$ and $f_2(\dums_2, \dums_3)$   are  defined in \eqref{eq:f1DDreg} and \eqref{eq:f2DDreg}.
		Moreover, the number of configurations with $\ldeg j$ type-$1$ edges,  $b_2 \rdeg -(i+j+k) \ldeg +b_1$ type-$2$ edges, $b_1 (\rdeg-1)$ type-$3$ edges, $k \ldeg $ type-$4$ edges, $ i \ldeg -b_1$ type-$5$ edges, $b_1$ type-$6$ edges such that there are exactly $b_1$ positive right nodes each one of them is connected to  exactly  one type-$6$ edge and $\rdeg-1$ type-$3$ edges, and $b_2$ positive right nodes each one of them is connected to at least one edge of type-$1$ or $5$ and it cannot be connected to one type-$1$ and $\rdeg -1$ type-$2$ or one type-$5$ and $\rdeg -1$ type-$2$ edges, is 
		\begin{equation}\label{eq:betaDDreg}
			\begin{aligned}
				&	{m \choose b_1,b_2,m-b_1-b_2} \rdeg^{b_1} \times \\& \coeff{g(\bttx)^{b_2},\dumx_1^{b_2 \rdeg+b_1-(i+j+k)\ldeg}\dumx_2^{k \ldeg} \dumx_3^{i \ldeg -b_1}}
			\end{aligned}
		\end{equation}
		where $g(\bttx)$ is defined in \eqref{eq:gDDreg}. 
		The proof follows by multiplying \eqref{eq:alphaDDreg} with \eqref{eq:betaDDreg}, dividing the result by the total number of configurations for the given distribution of edge types, and 
		summing over all possible $b_1, b_2$ and $k$.
	\end{proof}
\end{lemma}

\begin{figure*}[t]
	\begin{align}
		A_{i+j,j}^{\DD}=&  \sum\limits_{w,e,l, u,o,b_1} \frac{  \coeff{g(\bttx)^n,\dumx_1^{w}\dumx_2^{e} \dumx_3^{l}\dumx_4^{u}\dumx_5^{o} \dumx_6^{b_1}}   \coeff{f(\bttt,\btts)^{n},\dumt_1^i   \dumt_2^j    \dums_1^w \dums_2^e \dums_3^l \dums_4^{u}\dums_5^{o} \dums_6^{b_1}}}{{n{\bldeg} \choose w,e,l,u,o,b_1,n \bldeg-w-e-l-u-o-b_1}}\label{eq:AijDDireg}\\
		g(\bttx)&=\prod\limits_{d=1}^{\rdeg_{\text{max}}} \left[ 1+ (\dumx_1+\dumx_2+\dumx_4+\dumx_5)^d- (\dumx_2+\dumx_4)^d - d \dumx_1 \dumx_2^{d-1} - d \dumx_5 \dumx_2^{d-1}+d \dumx_3^{d-1} \dumx_6 \right]^{\xi \RH_d} \label{eq:gDDirreg}\\
		f(\bttt,\btts)&=\prod\limits_{d=1}^{\ldeg_{\text{max}}} \left[ \dumt_1 \left((\dums_5+\dums_6)^d)-\dums_5^d \right)+ \dumt_2 \dums_1^d+\dums_4^d+ (1+\dums_2+\dums_3)^d-(\dums_2+ \dums_3)^d \right]^{ \LA_d} \label{eq:fDDirreg}
	\end{align}
	\hrulefill
\end{figure*}

\begin{lemma}[Average enumerator for irregular ensembles]
	Under \ac{DD} detection, the average number of patterns with $i+j$ defective elements and $j$ misdetections in a pooling graph drawn uniformly at random from the ensemble $\Cirreg$ is given in \eqref{eq:AijDDireg} at the top of the next page, where $\bttx=(\dumx_1,\dumx_2,\dumx_3,\dumx_4,\dumx_5,\dumx_6)$, $\bttt=(\dumt_1,\dumt_2)$, $\btts=(\dums_1,\dums_2,\dums_3,\dums_4,\dums_5,\dums_6)$, $g(\bttx)$ and $f(\bttt,\btts)$ are given in \eqref{eq:gDDirreg} and \eqref{eq:fDDirreg} and where the sum is over all integers $w,e,l, u,o, b_1$ such that
	$l \leq (n-i-j-k) \ldeg_{\text{max}}$, $
	w+e+u+o \leq b_2 \rdeg_{\text{max}}$, $
	n-i-j-k \leq (m-b_1-b_2) \rdeg_{\text{max}}$, $
	b_1+b_2 \leq  (i+j) \ldeg_{\text{max}}$, and $
	i \leq b_1 \leq  i \ldeg_{\text{max}}$.
	\begin{proof}
		Consider a pooling graph drawn uniformly at random from the ensemble $\Cirreg$. We randomly chose a set $\defset$ of $i+j$ left nodes ($\defset$ is the set of defectives). 
		Here again, we have two types of positive right nodes. Positive right nodes of the first type are connected to exactly one \ac{PD} left node (which will be detected as $(1,1)$). The remaining positive right nodes are positive right nodes of the second type. We denote by $b_1$ the number of type-$1$ positive right nodes and $b_2$ the number of type-$2$ positive right nodes.
		We have six types of edges emanating from positive right nodes. Edges of the first type are connected to left nodes with state $(1,0)$ and type-$2$ positive right nodes, edges of the second type are connected to \ac{DND} left nodes and type-$2$ positive right nodes, edges of the third type are connected to \ac{DND} left nodes and type-$1$ positive right nodes, edges of the fourth type are connected to negative left nodes that are not connected to any negative right nodes and type-$2$ positive right nodes, edges of the fifth type are connected to \ac{DD}  left nodes and type-$2$ positive right nodes and edges of the sixth type are connected to \ac{DD}  left nodes and type-$1$ positive right nodes. We denote by $\alpha(i,j,w,e,l,u,o,b_1)$ the  number of ways to choose $i+j$ defective left nodes  such that we have  exactly $j$ left nodes with state $(1,0)$,  $w$ type-$1$ edges, $e$ type-$2$ edges, $l$ type-$3$ edges, $u$ type-$4$ edges, $o$ type-$5$ edges and $b_1$ type-$6$ edges. Its generating function is given by
		\[
		\sum_{i,j,w,e,l,u,o,b_1} \alpha(i,j,w,e,l,u,o,b_1) \dumt_1^i   \dumt_2^j    \dums_1^w \dums_2^e \dums_3^l \dums_4^{u}\dums_5^{o} \dums_6^{b_1}.
		\]
		Consider a single left node of degree $d$. If it has the state $(1,1)$, then we obtain $a \geq 1$ type-$6$ edges and  $d-a$ type-$5$ edges. If it has the state $(1,0)$, then we obtain  $d$ type-$1$ edges. If it has state $(0,0)$ and it is not connected to any negative right nodes, we obtain  $d$ type-$4$ edges, and if it has state $(0,0)$ and it is connected to at least one negative test, then we obtain at most $d-1$ type-$2$ and $3$ edges. Considering all possible left node degrees and since we have $n \LA_d$ degree $d$ left nodes, we obtain
		\begin{equation}\label{eq:alphaDDirr}
			\begin{aligned}			\alpha(i,j,w,&e,l,u,o,b_1)=\\ &\coeff{f(\bttt,\btts)^n,\dumt_1^i   \dumt_2^j    \dums_1^w \dums_2^e \dums_3^l \dums_4^{u}\dums_5^{o} \dums_6^{b_1}}.
			\end{aligned}
		\end{equation}
		Let $\beta(w,e,l,u,o,b_1)$ be the number of ways to choose  $w$ type-$1$ edges, $e$ type-$2$ edges, $l$ type-$3$ edges, $u$ type-$4$ edges, $o$ type-$5$ edges, $b_1$ type-$6$ edges. Its generating function is
		\[
		\sum_{w,e,l,u,o,b_1} \beta(w,e,l,u,o,b_1) \dumx_1^{w}\dumx_2^{e} \dumx_3^{l}\dumx_4^{u}\dumx_5^{o} \dumx_6^{b_1}.
		\]
		Consider a single right node of degree $d$. If it is negative then we obtain $0$ type-$1$-$6$ edges.  If it is positive and of type-$1$, then we obtain exactly $1$ type-$6$ edge and $d-1$ type-$3$ edges. If it is positive and of type-$2$, then we obtain at least one edge of type-$1$ or $5$, and it cannot be connected to one type-$1$ and $d -1$ type-$2$ or one type-$5$ and $d -1$ type-$2$.
		Considering all possible right node degrees and since we have $m  \RH_d$ degree $d$ right nodes, we obtain 
		\begin{equation}\label{eq:betaDDirr}
			\begin{aligned}
				\beta(w,e,l,u,o,b_1)= \coeff{g(\bttx)^n ,   \dumx_1^{w}\dumx_2^{e} \dumx_3^{l}\dumx_4^{u}\dumx_5^{o} \dumx_6^{b_1}}.
			\end{aligned}
		\end{equation}
		The proof follows by multiplying \eqref{eq:alphaDDirr} with \eqref{eq:betaDDirr}, dividing the result by the total number of configurations for the given distribution of edge types, and 
		summing over $b_1, b_2$ and $k$.        
		\par \vspace{-1\baselineskip}
		\qedhere
	\end{proof}
\end{lemma}

\subsection{Misdetection Probability}
The probability of having $j$ false negatives conditioned on having $i+j$ defective elements is
$Q(j|i+j)={n \choose i+j}^{-1}A_{i+j,j}^{\DD}$
and the \ac{MD}  probability for a population with prevalence $\delta$ is
\begin{align}
	\pfn^{\DD}=&\sum\limits_{a=1}^{n} \sum\limits_{j=1}^{a}\frac{j}{a} Q(j|a) {n \choose a}\delta^a(1-\delta)^{n-a} \\=&\sum\limits_{a=1}^{n} \sum\limits_{j=1}^{a}\frac{j}{a}A_{a,j}^{\DD} \delta^a(1-\delta)^{n-a}.\label{eq:pfn}
\end{align}

\section{Examples of Application}\label{sec:examples}

\begin{figure}[t]
	\centering
	\includegraphics[width=0.81\columnwidth]{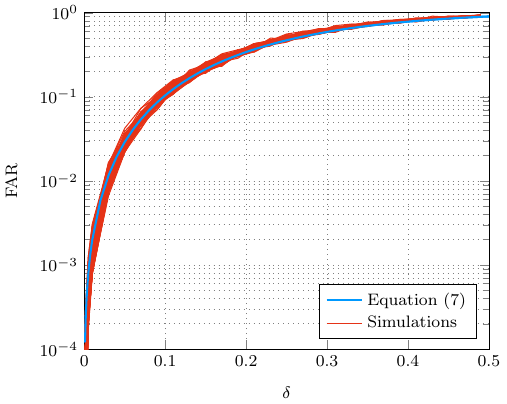}
	\vspace{-3mm}
	\caption{FAR  versus $\delta$ for graphs drawn from the  ensemble   $\Cregn{30}{3}{6}$ under COMP detection, compared with the ensemble average from   \eqref{eq:pfa}.}
	\label{fig:COMP_36_30}
\end{figure}

\begin{figure}[t]
	\centering
	\includegraphics[width=0.81\columnwidth]{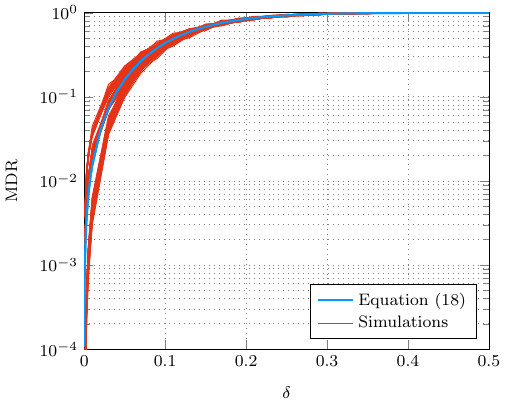}
	\vspace{-3mm}
	\caption{MDR versus $\delta$ for graphs drawn from the  ensemble  $\Cregn{30}{3}{6}$ under DD detection, compared with the ensemble average from  \eqref{eq:pfn}.}
	\label{fig:DD_36_30}
\end{figure}

Next, we provide examples applying the analysis of Section \ref{sec:COMP} and Section \ref{sec:DD}. For a given pooling graph ensemble, we compute the \ac{FA}/\ac{MD} probabilities of \ac{COMP}/\ac{DD} detection according to \eqref{eq:pfa} and \eqref{eq:pfn}, and we compare them with the \ac{FAR} and \ac{MDR} measured via Monte Carlo simulations, for $100$ realizations of random pooling graphs. We consider the $\Cregn{30}{3}{6}$ regular graph ensemble as a case study. The \ac{FAR} under \ac{COMP} is depicted as a function of $\delta$ in Figure \ref{fig:COMP_36_30}, while the \ac{MDR} under \ac{DD} detection is provided in Figure \ref{fig:DD_36_30}. The simulation results are closely concentrated around the ensemble averages \eqref{eq:pfa} and \eqref{eq:pfn}.  

\section{Conclusions}\label{sec:conc}
A combinatorial analysis of the probabilities of false alarm (FA) and misdetection (MD) of non-adaptive group testing with sparse pooling graphs was presented. The combinatorial orthogonal matching pursuit and definite defective detection algorithms were considered in the noiseless, non-quantitative setting. The analysis allows calculating the exact average FA/MD probabilities for pooling graph ensembles.

\end{document}